\documentclass{PoS}
\usepackage{epsfig}
\PoS{PoS(BDMH2004)076}
\title{Beyond the sphericity assumption in dynamical HI models}
\ShortTitle{Beyond the sphericity assumption in dynamical HI models}
\author{\speaker{Pieter Buyle}\\
        Ghent Univ., Belgium\\
        E-mail: \email{Pieter.Buyle@UGent.be}}
\author{Herwig Dejonghe\\
         Ghent Univ., Belgium\\ 
        E-mail: \email{Herwig.Dejonghe@UGent.be}}
\author{Gianfranco Gentile\\
         SISSA, Italy\\ 
        E-mail: \email{gentile@sissa.it}}

\abstract{It is generally agreed that the rotational velocities of
spiral galaxies indicate that a lot of matter is unseen. Hitherto, the
majority of these mass models are all based on spherical symmetry,
although this assumption is obviously not the one that is closest to
reality.  We present a new method to derive dynamical mass models from
HI-data based on an axisymmetric geometry. To validate our method, we
started a project with the aim to create such models from a combined
sample of late-type Low Surface Brightness galaxies and early-type
High Surface Brightness spiral galaxies. The whole sample spans a wide
range of bulge-to-disk ratios and will ultimately allow us to
investigate the orbital distribution as a function of morphology.}

\FullConference{Baryons in Dark Matter Halos\\
 5-9 October 2004\\
 Novigrad, Croatia}
\begin{document}
\section{Introduction}
Since the first observations of neutral hydrogen in our own galaxy
were made, dynamical HI models have only gained importance. Because
the radiation of HI is not affected by dust, it became the primary
tool to map the kinematics of galaxies, even beyond the optical
radius. This has lead to the dark matter paradigm, as the observations
showed that one had to invoke a lot of unseen matter to explain
the flattening of the rotation curves at high distances from the
centre. Although it is clear that this extra matter should be present
(in the Newtonian world view, that is), the amount and location may
also depend on the used dynamical mass modeling. The majority of
these models are based on spherical symmetry, which is probably not
the most suitable representation of the mass components in galaxies,
e.g. CDM simulations and observations suggest that the dark matter
haloes are triaxial (\cite{cole} \cite{dubinski} \cite{jing}
\cite{olling} \cite{sicking} \cite{marel} \cite{warren}). We aim to
improve on the modeling part by creating axisymmetric mass models of
our mixed sample of late-type Low Surface Brightness (LSB) galaxies
and early-type High Surface Brightness (HSB) spiral galaxies that were
recently observed at 21cm.
\section{Sample selection}
We compiled a sample of 12 spiral galaxies along the Hubble sequence
that span a wide range in bulge-to-disk ratios. For the late-type
galaxies we preferred LSB galaxies above HSBs since many
investigations have shown that LSBs are dark matter dominated and thus
form the ideal objects for this project. On the contrary, for the
early-type sample we prefer HSB galaxies rather than LSB galaxies,
since early-type LSB galaxies are very rare and contain even less
neutral hydrogen than their HSB counterparts making them much more
difficult to observe at 21cm. The LSB galaxies were selected from the
LSB catalogues from Morshidi-Esslinger (1999) and Monnier-Ragaigne
(2003) on basis of their morphology, angular size, velocity width and
HI emission. The early-type galaxies are taken from the HSB catalogue
of D'onofrio et al. (1995) according to the same criteria.
\section{Observations}
On 26, 28, 29 May and 1 June 2004 we observed the 4 LSB galaxies
LSBGF300-026, NGC4965, IC4366 and IC2147 with the Australia Telescope
Compact Array (ATCA) in the 1.5B configuration. Because our spectral
line widths were smaller than 150\ km\ s$^{-1}$ we selected a
correlator setup which yielded a bandwidth of 4\ MHz divided over 1024
channels, hence giving a channelwidth of 3.9\ KHz. Each galaxy was
observed for 12h to cover the whole U-V plane. Standard data reduction
was performed in the MIRIAD software package. Final datacubes were
made with a velocity resolution of 3.3\ km\ s$^{-1}$, though for
IC4366 we performed Hanning smoothing an additional time (yielding
a velocity resolution of 6.6\ km\ s$^{-1}$) in order to increase the S/N
ratio. The properties of the final observations are listed in Table
1. The 8 other galaxies will be observed in December 2004 and January
2005.\\
\begin{table}
\begin{tabular}{ccccc}
\hline
Galaxy name & LSBF300-026 & NGC4965 & IC4366 & IC2147\\
\hline
RA & 03h09m37.8s & 13h07m09.4s & 14h05m11.5s & 05h43m28.0s\\
Dec & -41$^{\circ}$01'50'' & -28$^{\circ}$13'41'' & -33$^{\circ}$45'36'' & -30$^{\circ}$29'42''\\
\# channels & 512 & 512 & 512 & 512\\
Channel seperation (KHz) & 7.8 & 7.8 & 7.8 & 7.8\\
Synth. beam (arcsec) & 35$\ \times\ $22 & 47$\ \times\ $24 & 43$\ \times\ $23 & 43$\ \times\ $21\\
Synth. beam (kpc) & 2.2$\ \times\ $1.4 & 7.0$\ \times\ $3.5 &12.9$\ \times\ $7 &3.7$\ \times\ $1.8\\
Noise (mJy\ beam$^{-1}$) & 1.7 & 1.9 & 1.7 & 1.9\\
\hline
\end{tabular}
\caption{Observational parameters}
\end{table}
\begin{center}
\epsfysize=4.3 cm
\epsfbox{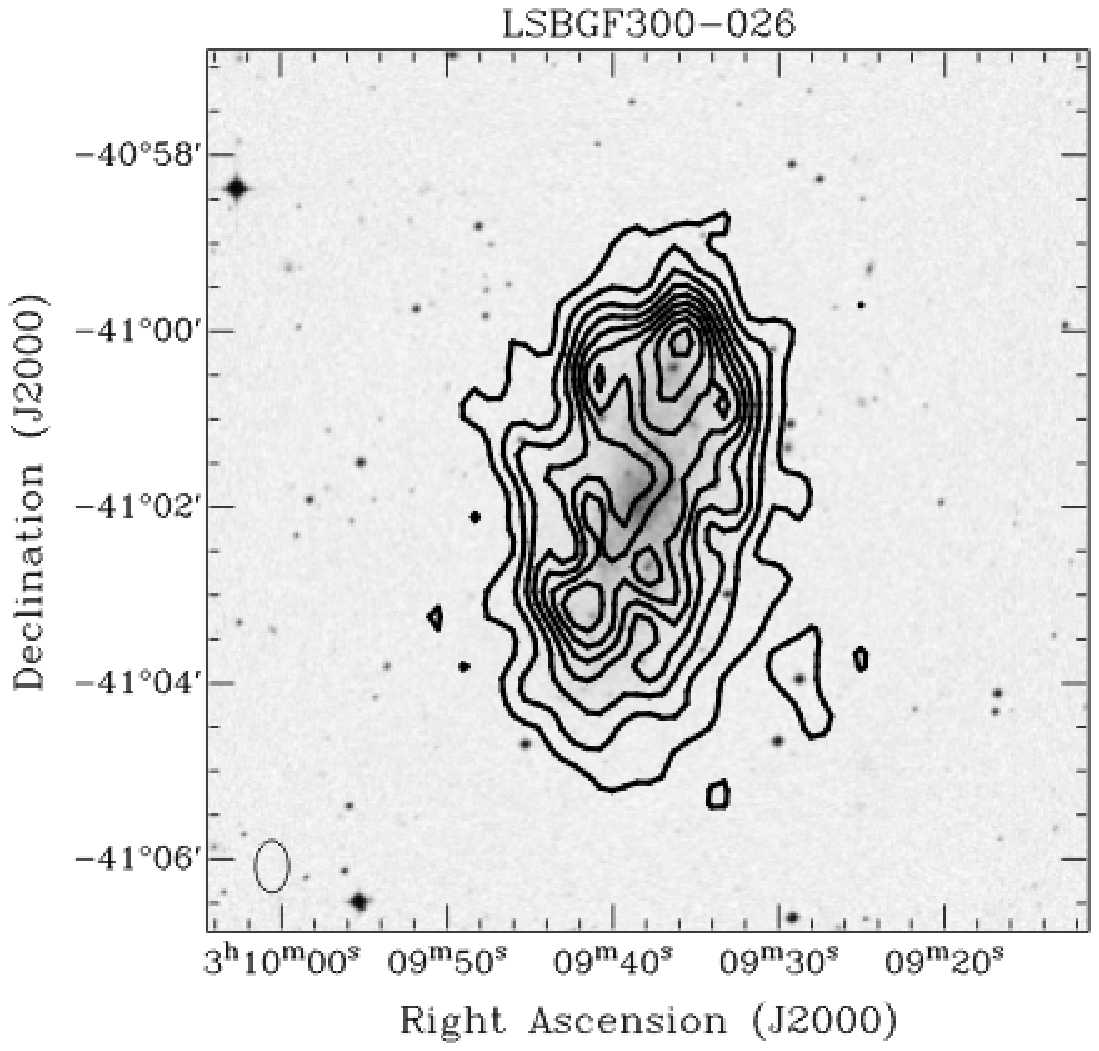}
\epsfysize=4.3 cm
\epsfbox{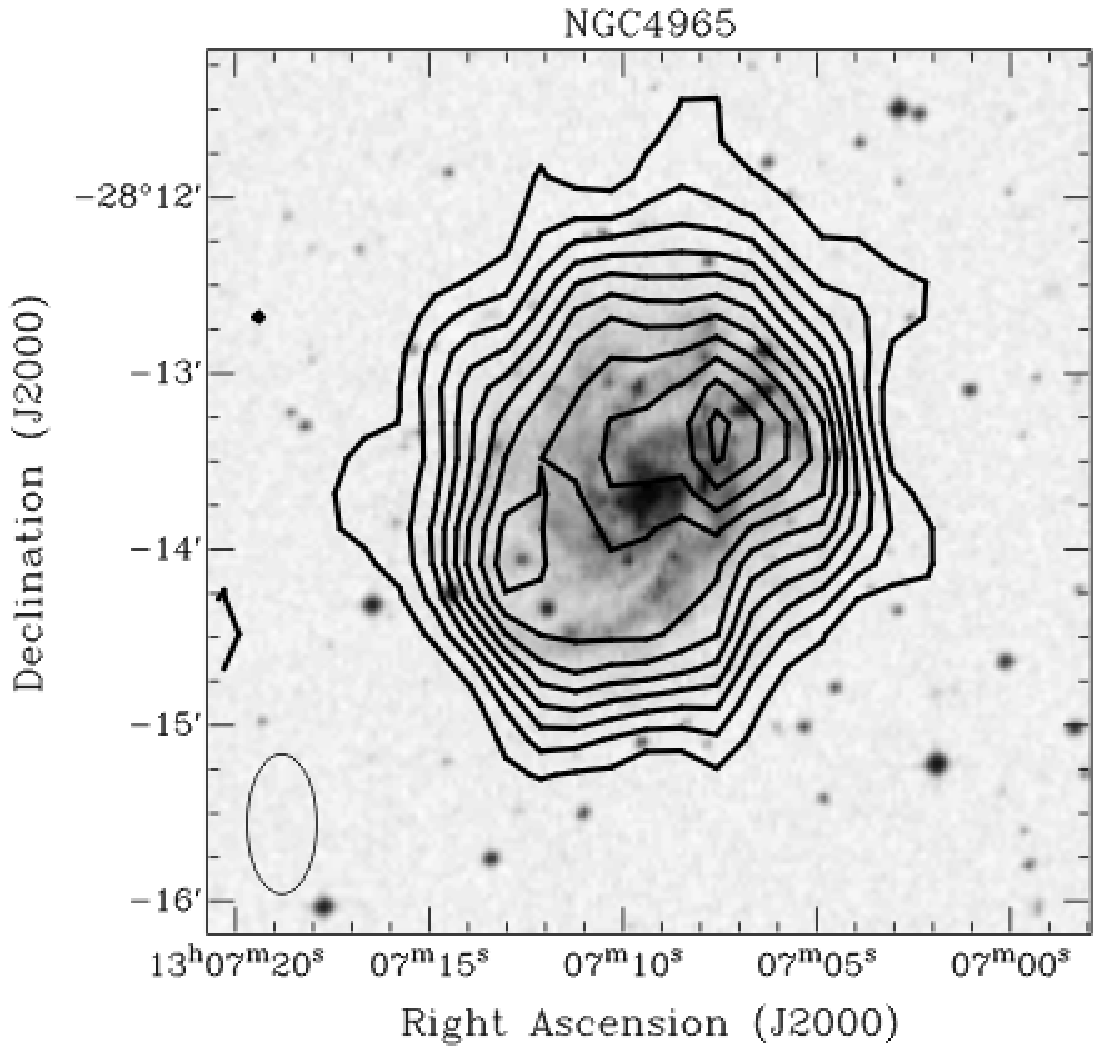}\\
\epsfysize=4.3 cm
\epsfbox{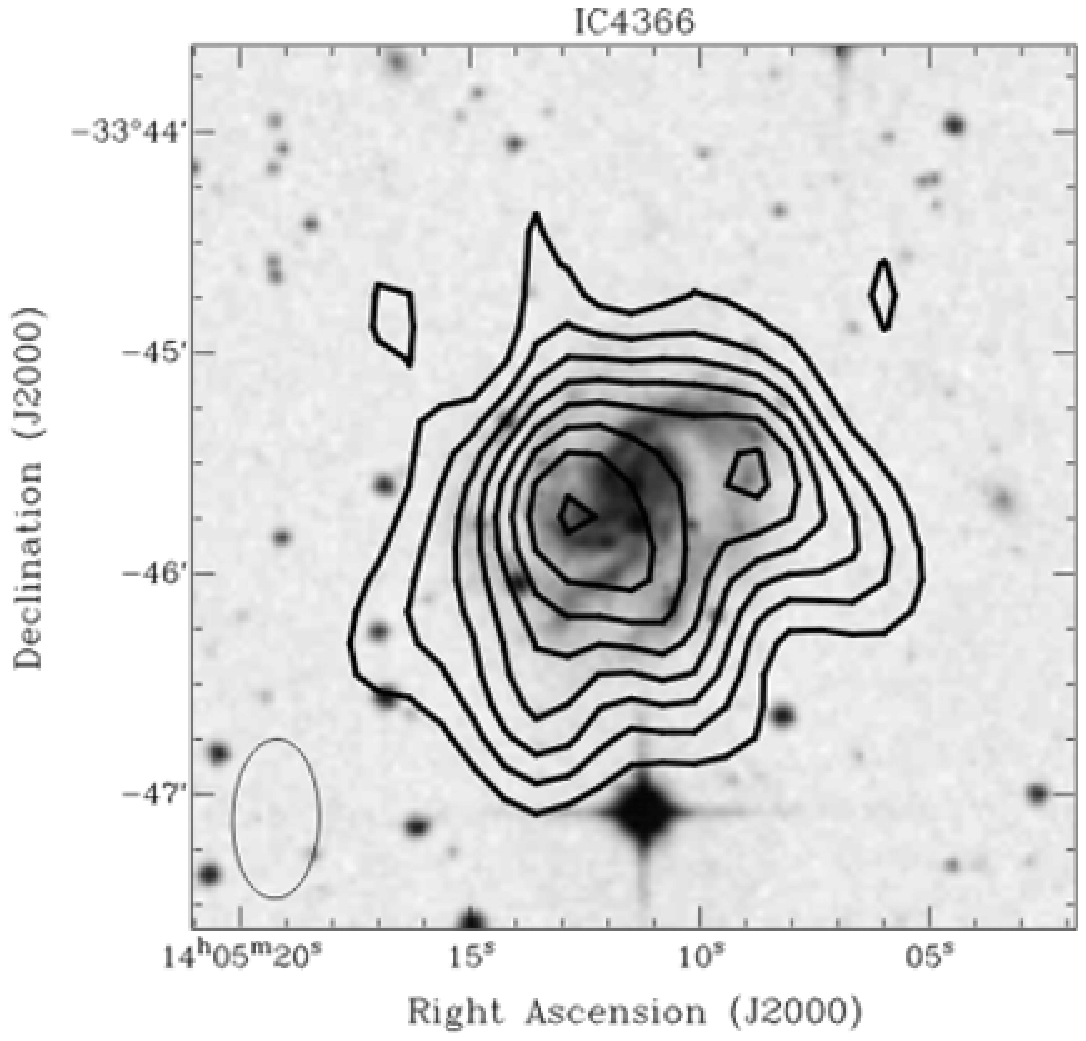}
\epsfysize=4.3 cm
\epsfbox{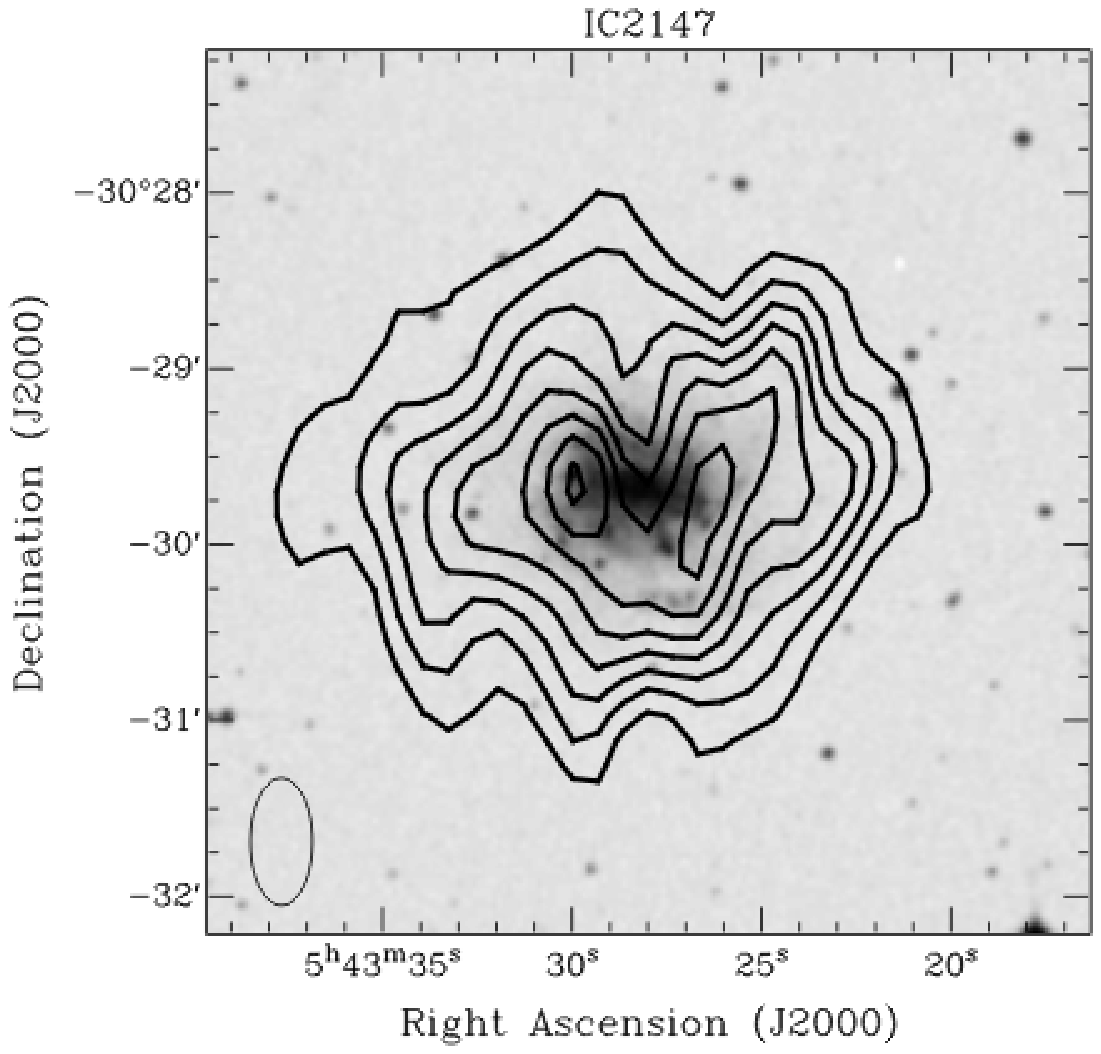}
\end{center}
{\small\em {\bf Figure 1}. The total-intensity maps (contours)
superimposed on optical DSS images (gray-scale). The outermost contour
corresponds to 3$\sigma$. The beam is plotted in the lower left
corner.}
\section{Axisymmetric models}
In order to construct axisymmetric models, one must assume the fact that
the data are compatible with an axisymmetric geometry. Hence, the
final datacubes we derived after datareduction were slightly modified
into an axisymmetric data set. An example of the result of this
process is shown in Figure 2, where it appears that axisymmetric
systems do in fact fit well to the observed data especially in the
outer regions. The inner regions are more distorted due to the effect
of spiral arms and star formation. Once a final axisymmetric data set
has been created, a density function must be obtained from the
total-intensity map. This will be done by quadratic programming
deprojection on a space of suitable basis functions for which the
gravitational potential can be calculated in a convenient
way\cite{dejonghe}. Additionally it is assumed that 1\ M$_{\odot}$ of
HI has an emissivity of
\begin{equation}
\epsilon=\frac{\textrm{h}\nu_0}{4\pi}\frac{3}{4}\textrm{N}_{\textrm{H}}\textrm{A}_{10}\phi(\nu)\ =\ 1.913\ \times\ 10^{17}\phi(\nu)\ \textrm{W}/\textrm{Hz}
\end{equation}
where $\phi(\nu)$ is a function of frequency and selfabsorption of the
HI-emission is neglected. This density - potential pair is usually
compatible with a vast number of distribution functions (which
distributes the gas and stars in a system). However, quadratic
programming allows us to $\chi^2$-fit the moments of the distribution
function to the observed kinematics derived from the full datacube.
\begin{center}
\epsfysize=5 cm
\epsfxsize=5 cm
\epsfbox{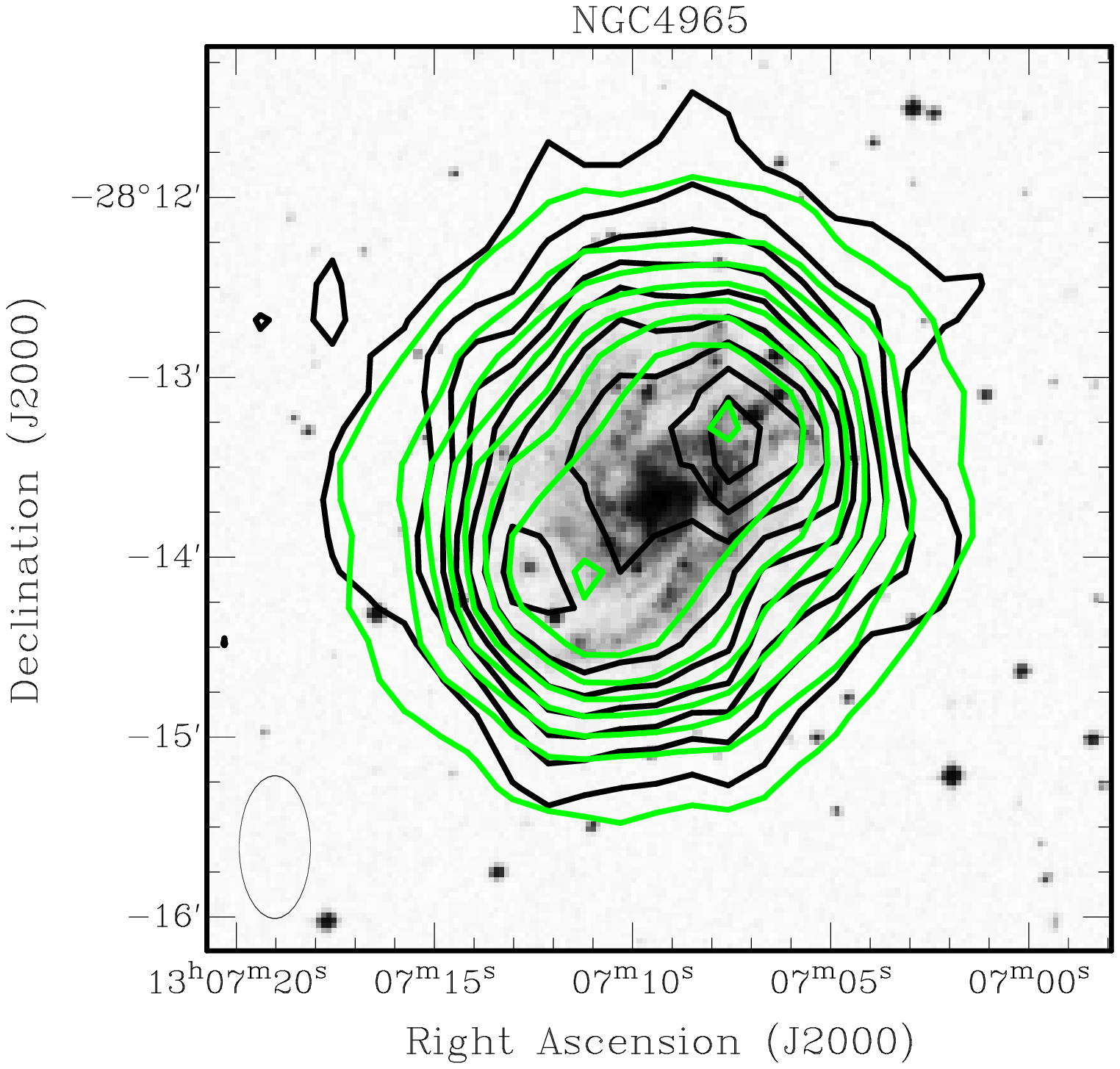}
\end{center}
{\small\em {\bf Figure 2}. The axisymmetric representation from the
datacube of NGC4965. The green contours are the model, while the black
shows the real datacube, superimposed on an optical image
(gray-scale). The contourlevels are the same for the model and
datacube.}
\section{Future}
The mass decomposition of the axisymmetric models relies on the
combination of HI-data and optical imaging. To trace the light of the
stars in our systems, we proposed deep H- and K-band imaging with the
NTT-telescope at ESO. When the final axisymmetric models of our
sample are constructed, we plan to investigate the
orbital distribution as a function of morphology. Additionally,
the models will enable us to examine the link between the putative
central supermassive black hole and dark matter which is suggested by the recently found $v_c$-$\sigma$ relation (\cite{ferrarese} \cite{baes}).

\end{document}